\documentstyle[12pt,aasms4]{article}  %%% This makes ms for submission

\slugcomment{Accepted, 1999 May 10th}

\lefthead{Belloni T. et al.}

\begin{document}

% our macros
\def\spose#1{\hbox to 0pt{#1\hss}}
\def\lta{\mathrel{\spose{\lower 3pt\hbox{$\mathchar"218$}}
     \raise 2.0pt\hbox{$\mathchar"13C$}}}
\def\gta{\mathrel{\spose{\lower 3pt\hbox{$\mathchar"218$}}
     \raise 2.0pt\hbox{$\mathchar"13E$}}}
\def\Msun{{\rm M}_\odot}
\def\msun{{\rm M}_\odot}
\def\Rsun{{\rm R}_\odot}
\def\Lsun{{\rm L}_\odot}
\def\half{{1\over2}}
\def\RL{R_{\rm L}}
\def\zs{\zeta_{s}}
\def\zR{\zeta_{\rm R}}
\def\dJJ{{\dot J\over J}}
\def\dMM{{\dot M_2\over M_2}}
\def\tKH{t_{\rm KH}}
\def\eck#1{\left\lbrack #1 \right\rbrack}
\def\rund#1{\left( #1 \right)}
\def\wave#1{\left\lbrace #1 \right\rbrace}
\def\dd{{\rm d}}

\title{A state transition of GX~339-4 observed with RXTE}
\author{
        T.~Belloni\altaffilmark{1,2},
        M.~M\'endez\altaffilmark{1,3},
        M.~van der Klis\altaffilmark{1},
        W.H.G. Lewin\altaffilmark{4},
        S. Dieters\altaffilmark{5}
       }

\altaffiltext{1} {Astronomical Institute ``Anton Pannekoek'',
       University of Amsterdam and Center for High-Energy Astrophysics,
       Kruislaan 403, NL-1098 SJ Amsterdam, the Netherlands}

\altaffiltext{2}{Present address: Osservatorio Astronomico di Brera,
	via Bianchi 46, I-23807 Merate, Italy}

\altaffiltext{3}{Facultad de Ciencias Astron\'omicas y Geof\'{\i}sicas, 
       Universidad Nacional de La Plata, Paseo del Bosque S/N, 
       1900 La Plata, Argentina}

\altaffiltext{4}{Massachussets Institute of Technology,
	         Center for Space Research, Room 37-627,
		 Cambridge, MA 02139, USA}

\altaffiltext{5}{CSPAR, University of Alabama in Huntsville, 
	Huntsville, AL 35899, USA}

\begin{abstract}
We report the results of observations with the Rossi X-ray Timing
Explorer of the black-hole candidate GX~339-4 during the state
transition of 1998. We find that both the X-ray spectrum  and the 
characteristic of the time variability after the transition are
typical of a high/soft state. We cannot exclude that the source went
through an Intermediate State before entering the high state. 
We discuss the results and compare them with other known black hole
candidates.

\end{abstract}

\keywords{accretion, accretion disks ---
          binaries: close  ---
          X-rays: stars --- stars: individual GX~339-4}

\section{INTRODUCTION}

GX 339-4 was discovered with the OSO-7 satellite (Markert et al. 1973),
and identified with a V$\sim$18 magnitude star in the optical (Doxsey et
al. 1979), with a $\sim$15 hour photometric period interpreted as the
orbital period (Callanan et al. 1992). Although no dynamical 
measurement of the mass of the compact object is available, the source
is included in the class of Black-Hole Candidates (BHCs) because of its
fast aperiodic variability and the occurrence of spectral/timing
transitions similar to those of established sources in the class (see
M\'endez \& van der Klis 1997). The source is usually observed in the 
Low State (LS), where the 1-10 keV energy spectrum 
is a power-law with spectral index
$\Gamma\sim$1.5-2.0 (Tananbaum et al. 1972) and the power spectrum
consists of a strong (25-50\% fractional rms) 
band-limited noise component similar
to that observed in Cyg X-1 (Oda et al. 1971, Miyamoto et al. 1992). In the High
State (HS), the source is brighter below 10 keV and an ultra-soft component
appears in the energy spectrum, while the power law component steepens
considerably; the power spectrum is reduced to a power law with a few percent
fractional rms (Grebenev et al. 1991).
In the Very-High State (VHS), observed on only one occasion, the source
is much brighter below 10 keV, mainly due to the increased luminosity
of the ultra-soft thermal component, the power law has a photon index of
$\Gamma\sim$2.5, and the power spectrum shows a 1-15\% rms variable
band-limited noise with a characteristic break frequency much higher than
in the LS (Miyamoto et al. 1991).
M\'endez \& van der Klis (1997) identified a fourth state in GX~339-4,
called Intermediate State (IS), observed also in GS~1124-68 by Belloni et 
al. (1997): its timing and spectral characteristics are similar to those
of the VHS, but the IS appears at much lower luminosities than the VHS. 
In GS~1124-68, as the outburst proceeded, the source moved from
the VHS to the HS to the IS and then to the LS, indicating that IS and VHS are 
indeed different states (see Belloni et al. 1997).
Currently, GX~339-4 is the only system, together with GS~1124-68, in which
all four states have been observed, although recently the superluminal
transient GRO~J1655-40 has shown a similar behavior (M\'endez, Belloni \& 
van der Klis (1998). 

In this letter, we report the results of RXTE/PCA observations of GX~339-4
during a transition to the high state in 1998 and compare them with previous
observations in the LS (analyzed in detail by Nowak, Wilms \& Dove 1999 and 
Wilms et al. 1999). %%In a companion paper, Fender et al. (1999) present
%%the results of radio observations covering the entire transition period.

\section{X-RAY OBSERVATIONS}

\subsection{All-Sky Monitor}

The All-Sky-Monitor (ASM: Levine et al. 1996) 
on board RXTE observed GX~339-4 almost continuously
since the beginning of the mission. The ASM light curve, in 1-day bins, is
shown in Fig. 1. The source was in a low-flux, hard state for the whole of
1996 and 1997. The flux level and the hardness ratio during this period
indicate that the source was in the low state. Some variability can be seen,
in the form of little ``outbursts'', whose ASM flux is
anti-correlated with the hard X-ray flux as observed by CGRO/BATSE
(see \cite{ral98}). In the beginning of
1998 January, a sharp increase in the ASM count rate is visible. The source
reached a level of approximately 20 cts/s and remained approximately constant
for $\sim$150 days before the flux started to decrease until it finally went
back to a low value (around $\sim$2 cts/s).
The switch to a higher count rate triggered a TOO observation
with the PCA/HEXTE instruments.

\subsection{PCA/HEXTE}

In response to the Target-of-Opportunity call, RXTE observed GX~339-4
for 45 ks between 1998 Jan 18 and Jan 15 (observation B, see Table 1). 
The time of the observations corresponds to the rise phase of the
outburst, just after a small flare (see Fig. 1). A month later the sources
reached the peak of the outburst (Fig. 1) and a second, much shorter, 
pointing was performed (observation C, see Table 1).
In addition, we analyzed a much older pointing extracted from the RXTE
archive (observation A, see Table 1), obtained when the source was in 
the LS (Fig. 1).

From PCA observations A and B, we divided the light curve from the
high-time resolution data in segments, produced 
a power spectrum from each segment, and averaged them together. The length
of each segment was 256s and 64s for observations A and B
respectively. All PCA channels were included in the analysis. We subtracted
from the resulting power spectra the contribution from the Poissonian noise
and the very large event window contribution (Zhang et al. 1995).
Because of its shortness, no useful data could be obtained from 
observation C. The two power spectra can be seen in Fig. 2.

Observation A looks like a typical LS, with a strong band-limited noise (BLN)
component. 
We fitted this power spectrum with a rather complicated model, consisting
of a broken power law,
a zero-centered Lorentzian of width $\nu_b$=0.75$\pm$0.04 Hz and a 
narrow QPO at $\nu_{1}$=0.35$\pm$0.03 Hz. In addition, a second QPO at
$\nu_2$=0.48$\pm$0.03 Hz (visible as a relatively small but significant 
feature in Fig. 2) and a broad Lorentzian bump at $\nu_3$=3.14$\pm$0.17
Hz were needed. An examination of more LS power spectra from other observations
showed that $\nu_1$ and $\nu_2$ are probably the second and third harmonics
of a fundamental $\nu_0\sim0.16$ Hz. The total fractional rms in the 0.01--100
Hz range is 41\%. The model used is different from that of the much
more complete work on the LS by Nowak, Wilms \& Dove (1999), but we use this
observation only for comparison. 
The power spectrum from observation B looks completely different. Not much
noise is observed and a simple power law model (with index 0.62$\pm$0.04)
gives a good fit to the data. The total rms in the 0.1--100 Hz band is
2\%. This weak noise component is characteristic of the HS. 

From all three observations, after checking that there were no large
flux variations, we extracted PCA and HEXTE energy spectra following the 
standard 
procedures for XTE, using {\tt ftools 4.2}. For spectral accumulation, 
we selected intervals where all 5 PCA detectors were turned on 
and the pointing offset was less than 0.02 degrees. 
%, not to complicate comparison between the two datasets. 
In order to minimize contamination, we further
selected data only from intervals when the Earth elevation angle
of the source was $>$10 degrees and the satellite
was well outside the South Atlantic Anomaly. PCA background files were 
produced with the program {\tt pcabackest} version 2.1b. We produced
PCA detector response matrices using {\tt pcarmf} v3.5.
For HEXTE, we accumulated background spectra from off-source pointings
and we used the latest background matrices made available by the RXTE team.
For observations B and C, not enough signal was present in the HEXTE
data above the first few channels, and those were 
therefore not used in the analysis. We used the HEXTE detector response
matrices from 1997 March 20th.  For the spectral fits, we used
{\tt XSPEC} 10.00 and added a 1\% systematic error to the PCA spectra.

We fitted the spectra with a rather complex but standard model, 
consisting of a 
power law with an exponential cutoff at high energies, a multicolor
disk blackbody (Mitsuda et al. 1984) 
and a gaussian emission line. Correction for interstellar
absorption was included, as well as a ``smeared edge'' (Ebisawa et al. 1994) 
which was found to be needed in order to obtain a satisfactory fit.
The central energy of the gaussian line was kept fixed at 6.4~keV.
Not all components were needed to model all spectra. No gaussian line
and no disk-blackbody components were needed for observation A, and no
smeared edge was needed for observation C (in this case,
because of the short exposure time).
Notice that, for the HS spectra, the emission line and the edge could arise
from the fact that the  continuum model is an approximation, since here the
thermal component is strong. Indeed, both the line and the edge are 
located in the 
energy range where the two continuum components become comparable. We do
not attempt a physical interpretation of these features.
The best fit parameters can be found in Table 2. The spectra and the residuals
after model subtraction are shown in Fig. 3.
It is evident from Table 2 that the large increase in X-ray flux between
observations B/C and observation A is due entirely to the appearance of
a soft thermal component, while the power law component steepens and 
becomes fainter.

\section{Discussion}

The RXTE/ASM light curve shown in Fig.~1 strongly suggests that GX~339-4
underwent a transition from the LS to the HS. The long-term behavior
%%is very similar to that observed in 1996 from Cyg~X-1 (Cui et al. 1997a,b),
consists basically of an interval of $\sim$400 days of increased ASM
flux. 
%(much longer than in the case of Cyg~X-1). 
Our power spectra and energy spectra are unambiguous: during observation 
A (reported also by other authors: Nowak, Wilms \& Dove 1999, Wilms et al. 1999)
the source was in its Low State, the state in which GX~339-4
is mostly observed. This is characterized by a flat ($\Gamma\sim$1.6) 
power-law energy spectrum (with evidence of a high-energy cutoff) and by
a strong band-limited noise in the power spectrum with a QPO peak and its
harmonics. At 4 kpc (see Zdziarski et al. 1998), 
the 2.5-20 keV observed luminosity is $\sim 4\times 10^{36}$erg/s.
Both energy 
distribution and power spectrum are extremely similar to those of Cyg~X-1.
This similarity includes the $\sim$3 Hz broad 
bump detected in the power spectrum.
Notice that the low-frequency QPO at 0.35 Hz and the 3~Hz bump have been 
shown by Psaltis, Belloni \& van der Klis (1999) to fit a correlation
which is observed when combining QPO data from a number of sources, both
containing neutron stars and black-hole candidates.

The X-ray properties of the source changed drastically after the
transition (observations B and C). Very little variability is observed
in the timing domain: the power spectrum shows only a weak power-law 
component. The energy spectrum is dominated by a thermal component, which
we fitted with the standard model used for black-hole candidates, i.e.
a multicolor disk-blackbody: the output parameters are temperature
of the inner edge of the accretion disk and the radius of the inner edge
itself. Interestingly, the radius which we derive is in the range 
expected for the inner-most stable orbit around a black hole, although 
the precise value cannot be determined as we do not know the inclination
of the system. Moreover, note that due to the approximated form of the 
disk-blackbody model used (see Mitsuda et al. 1984), the derived radius 
is likely to be smaller than the real one as the effective blackbody
temperature is probably smaller than the observed color temperature (
Lewin, van Paradijs \& Taam 1995).
The 2.5-20 keV luminosity of this component (at a distance
of 4 kpc) is 5$\times 10^{36}$ and 8$\times 10^{36}$erg/s for observation 
B and C, respectively,
whereas the corresponding luminosity of the (steeper) power law 
component is 10 and 4$\times 10^{35}$erg/s, respectively.
The difference between the two pointings
indicates a further anticorrelation between the two components.
These parameters are very similar to what is observed for the HS both in 
GX~339-4 itself and in other
sources (GX~339-4: Grebenev et al. 1991; GS~1124-68: Ebisawa et al. 1994;
GRO~J1655-40: M\'endez, Belloni \& van der Klis; 4U~1630-47: Kuulkers et al.
1998; Oosterbroek et al. 1998; LMC~X-3, Ebisawa et al. 1993).
Interestingly, 
Fender et al. (1999) found an anticorrelation between the X-ray flux (from 
RXTE/ASM) and radio flux. They show that the radio flux is strongly
suppressed during the HS period. This is analogous to what is observed
in Cyg X-1, where there is a suppression of radio flux during transitions
to and from a LS (see Zhang et al. 1997b).
The transition is clearly detected by CGRO/BATSE, in the form 
of a anti-correlation with the ASM data (Fender et al. 1999).

Comparing our results with those of Cui et al. (1997a,b), we
can see that, despite the similarity in long-term light curves, 
in Cyg X-1 the situation is different. During the transition, a
bright soft component appears in the energy spectrum of Cyg~X-1, 
but the power law 
component remains relatively strong. Moreover, the power spectra show
either a band-limited noise component or a power law component, but always
with a fractional rms well above 10\%. This is also evident in the light 
curve in Fig. 1 from Cui et al. (1997b), where large variations can be seen.
Following Belloni et al. (1996), comparing our results for GX~339-4, we
confirm that Cyg X-1 during the transition of 1996 never reached the HS
(as observed in other sources like LMC X-3, always seen in this state, 
Ebisawa et al. 1993), but switched from the LS to the IS and back. 
However, there is no sign of the IS in our observations. 
M\'endez \& van der Klis (1997) compiled a list of flux thresholds for the
various states based on previous state transitions. From their list,
assuming a typical HS energy spectrum, we estimate
that, based on the previous transitions, 
the IS should not have started until a count rate of $\sim$30 cts/s 
had been reached in the RXTE/ASM.
If GX~339-4 went into the IS before our observations, it did it
at a different flux level. As a direct comparison, the IS in M\'endez
\& van der Klis had a 2-10 keV flux of 
1.5$\times 10^{-9}$erg cm$^{-2}$s$^{-1}$, a factor of 2.3 less than what
we observe here.
Notice that just before our observation B, a small peak is visible
in Fig. 1, at 15-18 ASM cts/s. It is possible that during this time
the source went indeed through an IS, although we cannot confirm it.
This indicates that, if flux is a good tracer
of accretion rate, it is
not the only parameter governing these transitions. 
%%%a fact confirmed
%%%by recent observations of the transient 4U~1630-47 (Dieters et al., in
%%%preparation; Oosterbroek et al. 1998).

A simple classification in terms of four basic states with a
definite dependence on flux (see van der Klis 1995 for a review)
%although it derives from the comparison of quite a few sources,
fails to reproduce the whole wealth of behaviors observed in black-hole
candidates.
In addition to sources which do not seem to follow 
the simple scheme outlined above (e.g. GRS~1915+105, Belloni 1998;
XTE~J1550-564, Sobczak et al. 1999; GS~2023+338, Zycki, Done \& Smith 1999),
there are other examples which indicate that
accretion rate is not the only parameter governing
these transitions. This is particularly clear in the case of the 1998 
outburst of 4U~1630-47,
where a transition between IS and HS was not followed by a reverse transition
as the source went back into quiescence
%%when, later in the outburst, the source flux fell well below the 
%%pre-transition level 
(Dieters et al. 1999 in preparation).
Moreover, in 1996, RXTE observed a state transition of Cyg X-1; 
the source increased
its soft X-ray flux by a factor of 3-4 (Cui 1996; Cui, Focke \& Swank
1996), while the bolometric
flux remained approximately constant (Zhang et al. 1997a). 

%%Despite its spectral and
%%timing characteristics, many authors interpreted the transient (a few months)
%%state as a HS (Cui et al. 1997a,b), while Belloni et al. (1996) 
%%identified it as an IS, casting
%%doubts on whether a transition to the HS had ever been observed in Cyg X-1.

The results presented in this paper, together with recent results for
4U~1630-47 and other transients like GRS~1915+105 (Belloni 1998), 
GRO~J1655-40 (M\'endez, Belloni \& van der Klis 1997; Tomsick et al. 1999),
XTE~J1550-564 (Cui et al. 1999; Sobczak et al. 1999) and XTE~J1748-288
(Revnivtsev, Trudolyubov \& Borozdin 1999; Focke \& Swank 1999) show that 
the classification in terms of 4 source states is followed faithfully
by some sources (like GRO~J1655-40, XTE~J1550-564 and XTE~J1748-288), 
but is complicated by the absence of a unique flux ``trigger'' for transitions 
between states (like in 4U~1630-47 and GX~339-4) and by a completely
different behavior in the case of GRS~1915+105.

\acknowledgements

MM is a fellow of the Consejo Nacional de Investigaciones
Cient\'{\i}ficas y T\'ecnicas de la Rep\'ublica Argentina.
This work was supported in part by % the Netherlands Organization for
%%%Scientific Research (NWO) under grant PGS 78-277 and 
the Netherlands
Foundation for Research in Astronomy (ASTRON) under grant 781-76-017.
TB is supported by NWO Spinoza grant 08-0 to E.P.J. van den Heuvel.
SD is supported by NASA LTSA grant NAG 5-6021.
WHGL acknowledges support from NASA.
This research has made use of data obtained through the High Energy 
Astrophysics Science Archive Research Center Online Service, provided
by the NASA/Goddard Space Flight Center.

{}

\clearpage

\figcaption[]{RXTE/ASM light curve of GX~339-4. Bin size is 1 day. 
	The arrows and the dotted lines indicate the times of the three 
	pointed observations discussed in the text. 
            }

\figcaption[]{Power spectra from observation A (filled circles) and B
	(open circles). 
            }

\figcaption[]{Energy spectra from observation A (top panel) and B/C (bottom
	panel). The residuals after subtraction of the best fit models
	described in the text are shown.
            }

\clearpage 

\begin{deluxetable}{lccc}
\footnotesize
\tablecaption{PCA/HEXTE observation log}
\tablewidth{0pt}
\tablehead{
  \colhead{Observation} &
  \colhead{Start}  &
  \colhead{End}  &
  \colhead{PCA exp. (s)} 
  \nl
}
\startdata
A & 1996--07--26 18:20UT & 1996--07--26 20:15UT & 5200\nl
B & 1998--01--15 03:56UT & 1998--01--18 03:20UT & 45400\nl
C & 1998--02--21 15:04UT & 1998--02--21 15:36UT & 1600\nl
\enddata
\end{deluxetable}
\clearpage 

\begin{deluxetable}{lccc}
\footnotesize
\tablecaption{Best-fit parameters for the energy spectra from the three
	observations of GX~339-4 with the model described in the text.
	Errors are 1$\sigma$. Fluxes are in the 2.5-20.0 keV band. 
        The inner disk radius is computed assuming a distance of 4 kpc.}
\tablewidth{0pt}
\tablehead{
  \colhead{Par.} &
  \colhead{Obs. A} &
  \colhead{Obs. B} &
  \colhead{Obs. C} \nl
}
\startdata

N$_H$ [cm$^{-2}$]      & (1.1$\pm$0.1)$\times 10^{21}$&
                         $<$9.7$\times 10^{21}$       &
                         $<$2.0$\times 10^{21}$       \nl

\hline %------------------------------------------------------------------

$\Gamma$               & 1.63$\pm$0.01 & 2.57$\pm$0.05 & 2.12$\pm$0.19 \nl

E$_{cut}$ [keV]        & 270$\pm$60    &  ---          &  ---          \nl

F$_{pl}$ [erg cm$^{-2}$s$^{-1}$]& 2.3$\times 10^{-9}$ & 
                                                 6.2$\times 10^{-10}$ &
                                                 2.0$\times 10^{-10}$ \nl

\hline %------------------------------------------------------------------

kT [keV]               & --- & 0.63$\pm$0.01   & 0.72$\pm$0.01 \nl
R$_{in}\sqrt{\cos i}$ [km]          & --- & 28.7$\pm$4.0    & 25.4$\pm$0.6  \nl
F$_{dbb}$ [erg cm$^{-2}$s$^{-1}$]& ---                 & 
                                                 2.5$\times 10^{-9}$ &
                                                 4.6$\times 10^{-9}$ \nl

\hline %------------------------------------------------------------------

E$_{lin}$ [keV]        & --- &  6.4 (FIX) & 6.4 (FIX) \nl
$\sigma_{lin}$ [keV]   & --- &  1.45$\pm$0.14 & 0.79$\pm$0.10 \nl
EW$_{lin}$ [eV]        & --- &  350           &  319          \nl

\hline %------------------------------------------------------------------

E$_{edg}$ [keV]        & 7.26$\pm$0.24   &  6.69$\pm$0.15 & --- \nl
$\tau_{edg}$           & 0.67$\pm$0.24   &  $>$ 6.2       & --- \nl
W$_{edg}$ [keV]        & 4.79$\pm$2.02   & 16.8$\pm$5.4   & --- \nl

\hline %------------------------------------------------------------------

$\chi_r^2$ (d.o.f.)    & 0.94  (146)     & 0.83 (41)      & 0.88 (46)\nl

\hline %------------------------------------------------------------------
\enddata
\end{deluxetable}

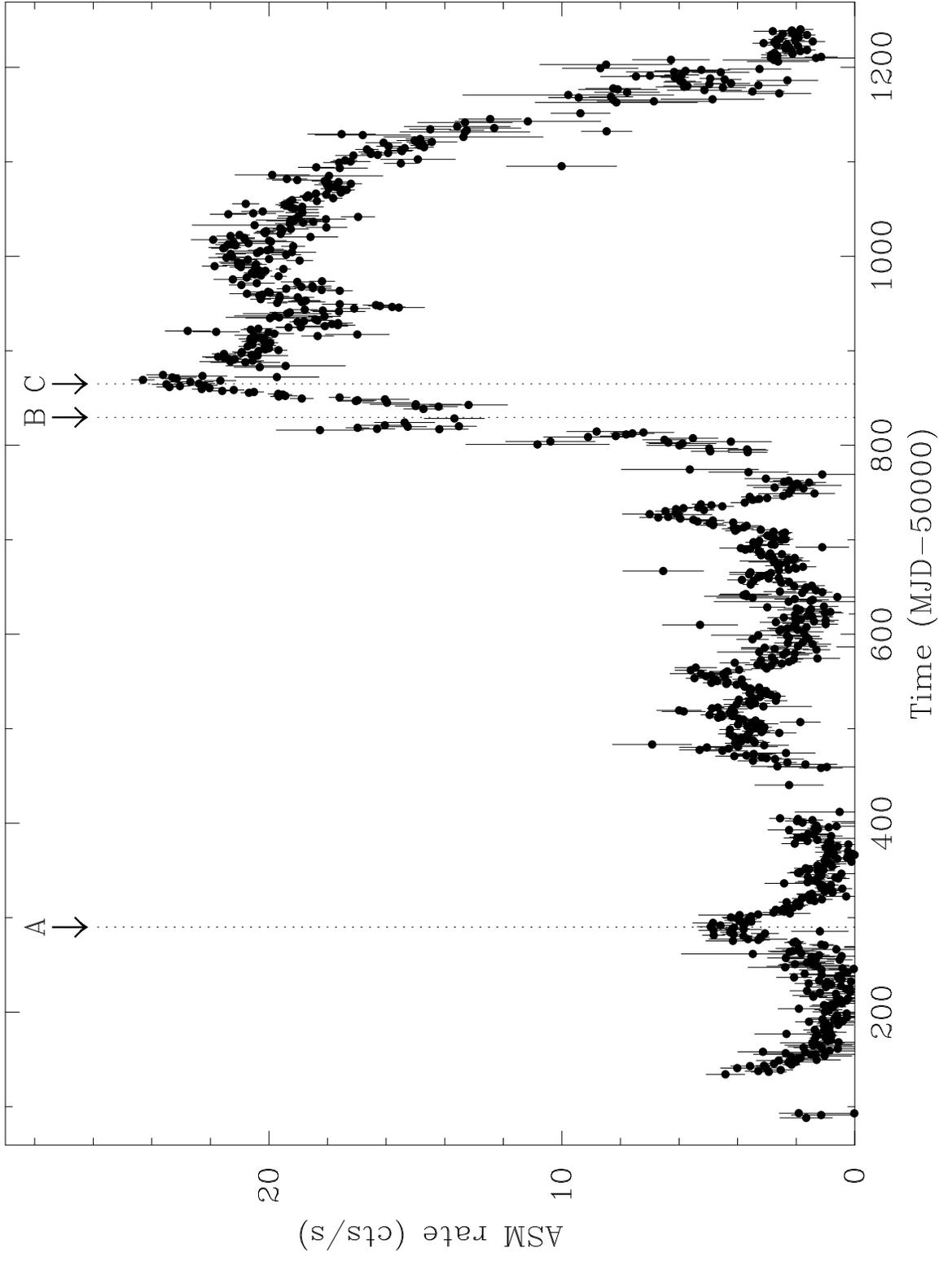
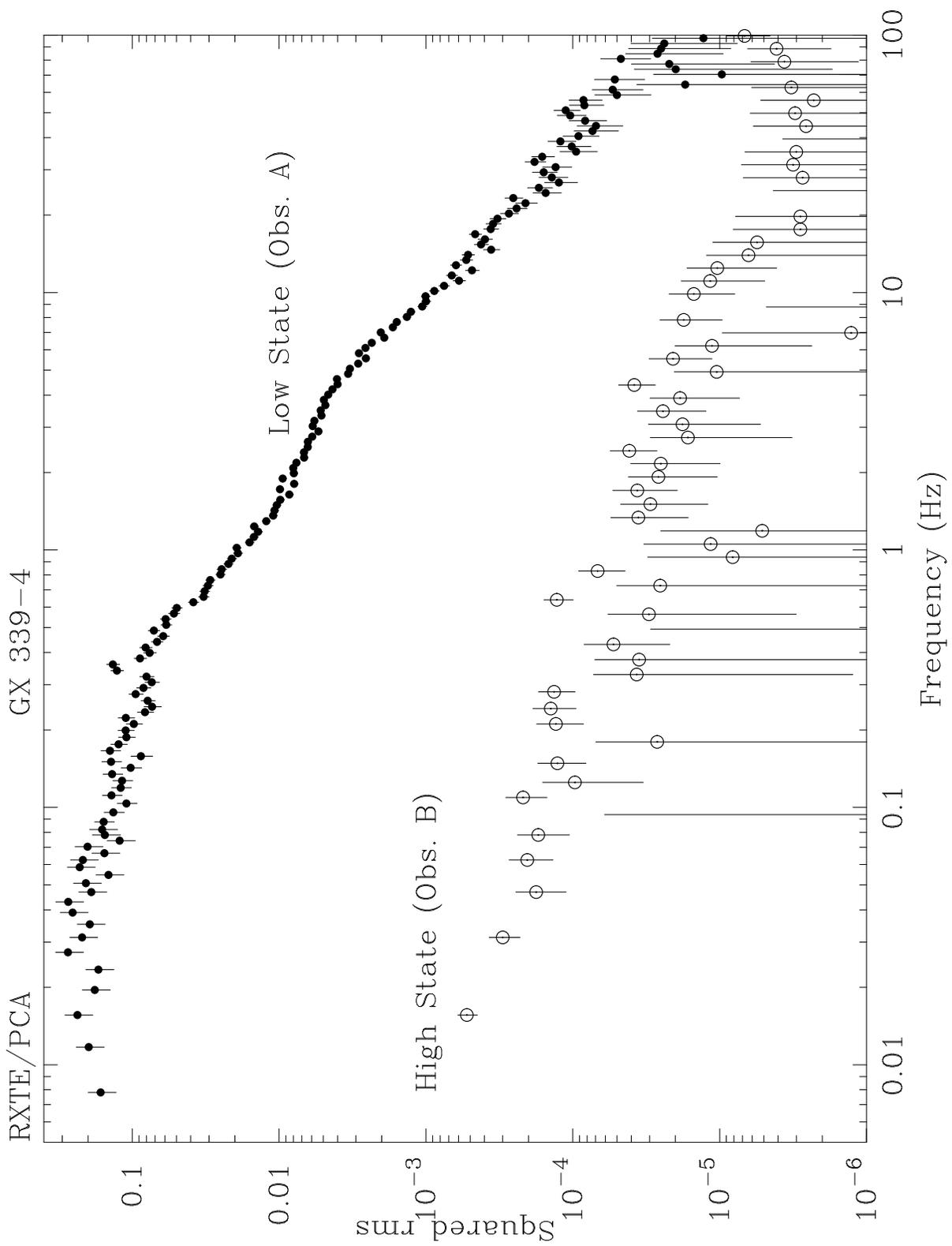
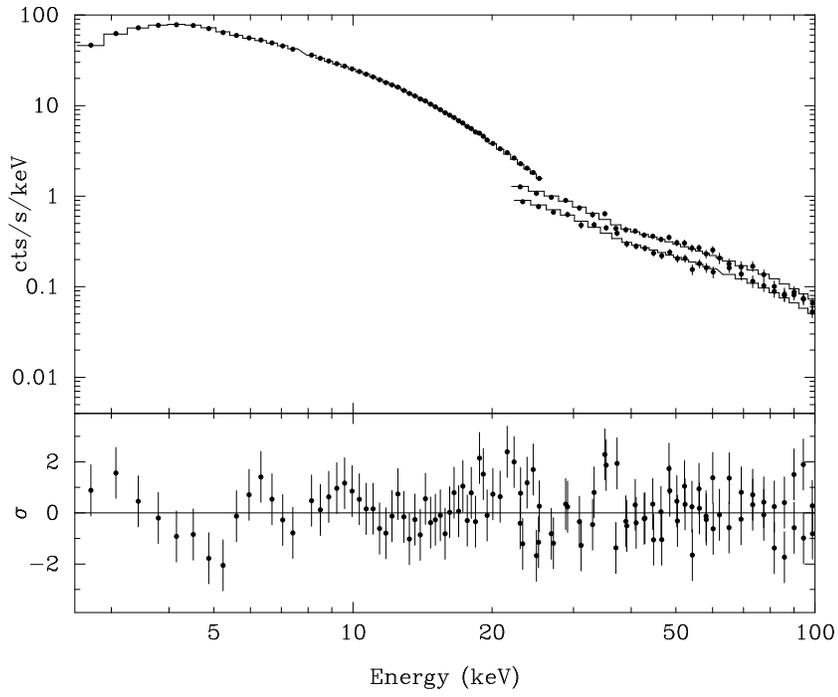
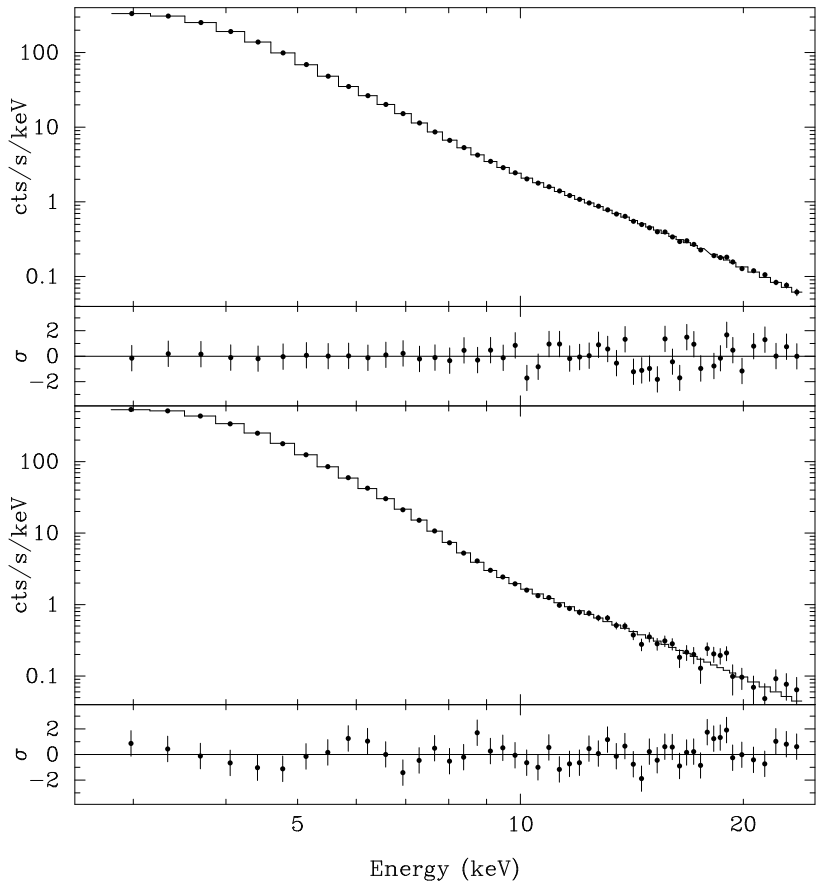
\end{document}